\documentclass[aip,amsmath,amssymb,reprint]{revtex4-1}
\newcommand{\ket}[1]{\left\vert #1 \right\rangle}
\newcommand{\bra}[1]{\left\langle #1 \right\vert}

\usepackage{graphicx}
\usepackage{epsfig}
\usepackage{amsmath}
\usepackage{amsfonts}
\usepackage{dsfont}
\usepackage{xcolor}
\usepackage{notes2bib}

\begin{document}
%
\bibliographystyle{apsrev}
\preprint{version: \today}
\title{A first principles derivation of energy conserving momentum jumps in surface hopping simulations}
\author{Dorothy Miaoyu Huang}
\author{Austin T. Green}
\author{Craig C. Martens}
\affiliation{University of California, Irvine, California 92697-2025}
\email[email: ]{cmartens@uci.edu}
\date{\today}
%
\begin{abstract}

The fewest switches surface hopping (FSSH) method proposed by Tully in 1990 [J. C Tully, J.~Chem.~Phys.~\textbf{93}, 1061 (1990)]---along with its many later variations---is basis for most practical simulations of molecular dynamics with electronic transitions in realistic systems.  Despite its popularity, a rigorous formal derivation of the algorithm has yet to be achieved.  In this paper, we derive the energy conserving momentum jumps characterizing FSSH from the perspective of quantum trajectory surface hopping (QTSH) [C. C. Martens, J.~Phys.~Chem. A \textbf{123}, 1110 (2019)].  In the limit of localized nonadiabatic transitions, simple mathematical and physical arguments allow the FSSH algorithm to be derived from first principles.  For general processes, the quantum forces characterizing the QTSH method provides accurate results for nonadiabatic dynamics with rigorous energy conservation at the ensemble level within the consistency of the underlying stochastic surface hopping without resorting to the artificial  momentum rescaling of FSSH.   

\end{abstract}
%
%
\maketitle
%


\section{Introduction}

The development and application of semiclassical and mixed quantum-classical methods for simulating molecular dynamics with nonadiabatic transitions using a classical trajectory framework is an important area of current research in theoretical chemistry. Current approaches include trajectory surface hopping, \cite{tull90,Subotnik:2016ch,Wang:2016bz} mapping Hamiltonians, \cite{bone01,Nassimi:2010ht,Nassimi:2009dn,rich2023} symmetrical windowing of quasiclassical trajectories, \cite{Cotton:2013fa,Cotton:2017ih} ring polymer methods, \cite{Richardson:2013jm,Ananth:2013fp} Gaussian wavepacket approaches, \cite{White:2016ic,Humeniuk:2016fj} exact factorization, \cite{Agostini:2014id,agocurch2019} quantum-classical Wigner function-based approaches, \cite{mart97,dono98,dono00,dono00b,kapr99,Kelly:2010dp,Bonella:2010hz,wan00a,wan02,ando02,ando03} to cite just a few.  In general, these methods hold great promise in that they allow the essential quantum features of complex systems to be incorporated within the computationally less demanding and intuitively appealing framework of classical mechanics.  

Currently, the most popular and practical quantum-classical trajectory method is fewest-switches surface hopping (FSSH), originally proposed by Tully in 1990 \cite{tull90}, and the subject of many later adaptations. \cite{Wang:2016bz,Subotnik:2016ch}  Here, individual trajectories undergo stochastic transitions from one (usually adiabatic) electronic state to another using probabilities calculated by solving a \emph{proxy} electronic Schr\"{o}dinger equation associated with each trajectory, which act as approximate local stand-ins for the exact electronic quantum subsystem, in parallel with the classical dynamics.  Thinking and computing with individual trajectories has many advantages, particularly in many-dimensional systems, where ``on the fly'' electronic structure methods are often employed to calculate forces and couplings at the instantaneous trajectory coordinates (see, e.g., \cite{benn02,Doltsinis:2002wt,Barbatti:2007cp,Curchod:2013dw,Tapavicza:2013bp,Zimmermann:2014ks,Persico:2014db}).  

Persistent challenges remain in correctly and consistently incorporating quantum effects in the manifestly classical framework of trajectory ensemble methods.  One serious outstanding problem is the correct treatment of quantum coherence and decoherence in the surface hopping algorithm.  In the original FSSH formulation,\cite{tull90,Tully:2012cz} the mixing of classical and quantum evolution at the independent trajectory level results in an approximation to the underlying quantum dynamics that is often too coherent, leading to significant errors in some applications.\cite{Subotnik:2016ch,Wang:2016bz} A range of attempts have been made to correct this over-coherence problem by adding dephasing to the method, either phenomenologically or based on other assumptions, with varying degrees of success. \cite{Jaeger:2012hp,BedardHearn:2005ib,Subotnik:2011bo,Shenvi:2012bk,Nelson:2013eh,Granucci:2010bh,Subotnik:2013jw}  

A second challenge involves the balancing of the quantum-classical energy budget during nonadiabatic processes.  Much thought and effort has gone into correcting the classical trajectory dynamics to incorporate the excess or deficit of quantum energy that accompanies electronic transitions.  The fundamental reason that problems arise is that surface hopping as conventionally formulated treats each trajectory as independent of the other ensemble members.  This greatly simplifies and speeds up the numerical implementation.  However, as we have long emphasized, \cite{dono98,dono00,dono00b,dono02b,riga04,riga06a,riga06b,roma04,roma07,dono01,dono03} important quantum mechanical effects arise in the trajectory framework through the interdependence of the members of the ensemble and a relaxation of strict classical conservation laws at the individual trajectory level.  

Tully's FSSH addresses the energy budget of quantum-classical molecular dynamics using a physically motivated but \emph{ad hoc} approach that imposes an external constraint of energy conservation at the individual trajectory level.  The constrained quantity is the classical kinetic plus potential energy of each trajectory.  For electronic transitions from higher to lower energy states, the conservation is imposed by accompanying the stochastic trajectory hop with an instantaneous classical momentum jump that increases the kinetic energy by the same amount as the decrease in electronic energy.  In practice, this is accomplished by solving a quadratic equation for the energy-conserving change in momentum.\cite{tull90,Subotnik:2016ch,Wang:2016bz}  Physical arguments suggest that the jump should be along the nonadiabatic coupling vector between the two involved adiabatic states.  The undetermined sign of the jump is assigned using physical arguments.\cite{Wang:2016bz,Subotnik:2016ch} 

For transitions from lower to higher electronic states, a similar procedure is employed to decrease the classical kinetic energy to accommodate the necessary increase in electronic energy.  If insufficient kinetic energy is available, the stochastic transition dictated by the quantum evolution and the stochastic algorithm is artificially prevented.  These aborted transitions are termed ``frustrated hops'', and are responsible for a breakdown of the consistency of the surface hopping method, defined as the agreement between the state occupancy statistics of the trajectory ensemble and the continuous populations of the proxy density matrices.\cite{tull90,Subotnik:2016ch,Wang:2016bz}

Attempts have been made to prove the validity of the FSSH algorithm from first principles, \cite{Subotnik:2013jw,kapr2016} but, in our opinion, no definitive derivation has yet been given.  

Recently, we proposed an alternative approach to trajectory surface hopping, which we call \emph{quantum trajectory surface hopping} (QTSH).\cite{martens2019surface,martens2020faraday}  QTSH is an independent trajectory approximation to consensus surface hopping (CSH), \cite{Martens:2016ea} our earlier method that solves the quantum-classical Liouville equations with interacting trajectories.  CSH does not introduce a proxy density matrix, as done in FSSH, but calculates the local quantum densities and coherences collectively from the trajectory ensemble.  QTSH, on the other hand, treats the trajectories independently, and employs the proxy density matrix approximations of FSSH.  

FSSH and QTSH differ significantly in their treatment of the quantum-classical energy budget.  In particular, QTSH employs quantum forces coupling the electronic transitions to the (now non-) classical trajectory evolution rather than employing the \emph{ad hoc} momentum jumps of FSSH.  These forces are derived from a rigorous solution of the coupled partial differential equations of the quantum-classical Liouville equation using a trajectory ensemble ansatz.\cite{martens2019surface,martens2020faraday} 

In this paper, we address the energy budget of quantum-classical dynamics from the perspective of independent trajectory surface hopping. Using the quantum-classical Liouville equation as a starting point, we briefly review the derivation of the quantum trajectory surface hopping (QTSH) method \cite{martens2019surface,martens2020faraday} and compare it with the existing FSSH approach.  By analyzing the effect of the quantum force of QTSH for localized nonadiabatic transitions, we derive the FSSH momentum jump in the limit of localized population transfer and show that classical energy conservation emerges rigorously from the interplay of the QTSH forces coupling electronic and nuclear degrees of freedom for general transitions.  

The organization of the rest of this paper is as follows.  In Sec.~\ref{sec:sec2}, we review the relevant background theory of nonadiabatic dynamics and the quantum-classical Liouville equation.  The quantum Hamiltonian and density matrix are briefly described for a system with two electronic states in the diabatic and adiabatic representations, as well as the semiclassical limit of the theory in the Wigner representation.  In this context, we consider a simple model of the density matrix evolution for localized transitions. Section \ref{sec:sec3} describes the QTSH method and compares its treatment of the quantum-classical energy budget with that of the FSSH approach. In Sec.~\ref{sec:sec4}, we analyze the quantum-classical energy budget and give a simple but rigorous derivation of the FSSH momentum jump from QTSH in the limit of localized transitions.  In Sec.~\ref{sec:sec5} we present numerical simulations of a nonadiabatic transition in a one-dimensional two state system.  We compare QTSH with exact quantum wavepacket results from the perspective provided by our analysis in Sec.~\ref{sec:sec4}.  A discussion is given in Sec.~\ref{sec:sec6}.

\section{Theory} \label{sec:sec2}

We start with the coupled electronic-nuclear dynamics of a two electronic state system in the diabatic representation.  The Hamiltonian is a $2 \times 2$ matrix of nuclear operators, given by
\begin{equation} \label{eq:hdia}
\hat{\mathbf{H}}_D = \hat{\mathbf{T}}_D + \mathbf{V}_D.
\end{equation}
The kinetic energy $\hat{\mathbf{T}}_D$ is diagonal in the diabatic representation:
\begin{equation} \label{eq:tdia}
\hat{\mathbf{T}}_D = \left( \begin{array}{lr} \hat{T}& 0 \\
0 &\hat{T} \end{array} \right),
\end{equation}
where $\hat{T}= -\frac{\hbar^2}{2m} \nabla^2$ is the nuclear kinetic energy operator. The potential matrix is non-diagonal in the diabatic representation, 
\begin{equation}
\mathbf{V}_D = \left( \begin{array}{lr} V_1(\mathbf{q})& V_{12}(\mathbf{q}) \\
V_{12}(\mathbf{q}) & V_2(\mathbf{q}) \end{array} \right),
\end{equation}
where $V_1(\mathbf{q})$ and $V_2(\mathbf{q})$ are the diabatic state potentials and $V_{12}(\mathbf{q})$ is the real valued off-diagonal diabatic coupling between electronic states $\ket{1}$ and $\ket{2}$.  

The potential matrix can be diagonalized by a unitary transformation $\mathbf{U}(\mathbf{q})$:
\begin{equation}
\mathbf{V}_A(\mathbf{q}) = \mathbf{U}^\dagger (\mathbf{q}) \mathbf{V}_D(\mathbf{q}) \mathbf{U}(\mathbf{q})=\left( \begin{array}{lr} V_+(\mathbf{q}) & 0 \\
0  &V_-(\mathbf{q}) \end{array} \right),
\end{equation}
where the unitary transformation matrix 
\begin{equation}
\mathbf{U}(\mathbf{q}) = \left( \begin{array}{lr} \cos(\frac{\phi(\mathbf{q})}{2})& -\sin(\frac{\phi(\mathbf{q})}{2}) \\
\sin(\frac{\phi(\mathbf{q})}{2})&\cos(\frac{\phi(\mathbf{q})}{2}) \end{array} \right)
\end{equation}
is a function of the coordinates $\mathbf{q}$ and
\begin{equation}
\phi(\mathbf{q}) = \tan^{-1}\left(\frac{2 V_{12}(\mathbf{q})}{V_1(\mathbf{q}) - V_2(\mathbf{q})}\right)
\end{equation}
is the mixing angle.  The coordinate dependence of $\phi(\mathbf{q})$ assures that the adiabatic potential matrix is diagonal for each configuration $\mathbf{q}$.

The eigenvalues of the matrix $\mathbf{V}_D$ are the adiabatic potentials, and are given by
\begin{equation}
V_{\pm}(\mathbf{q})=\frac{V_1(\mathbf{q})+V_2(\mathbf{q})}{2} \pm \sqrt{\left(\frac{V_1(\mathbf{q})-V_2(\mathbf{q})}{2}\right)^2 + V_{12}^2(\mathbf{q})}.
\end{equation}

Due to the coordinate dependence of $\mathbf{U}(\mathbf{q})$, the kinetic energy $\mathbf{T}_A = \mathbf{U}^\dagger (\mathbf{q}) \mathbf{T}_D \mathbf{U}(\mathbf{q})$ is \emph{not} diagonal in the adiabatic representation, but is given by
\begin{equation}
\mathbf{T}_A = \left( \begin{array}{ccc} \frac{1}{2 m} (\hat{\mathbf{p}}^2 + \hbar^2 \mathbf{d}^2)& \,\, & -\frac{i \hbar}{2 m}(\mathbf{d}\cdot \hat{\mathbf{p}} + \hat{\mathbf{p}} \cdot \mathbf{d}) \\
\frac{i \hbar}{2 m}(\mathbf{d}\cdot \hat{\mathbf{p}} + \hat{\mathbf{p}} \cdot \mathbf{d})  & \,\,& \frac{1}{2 m} (\hat{\mathbf{p}}^2 + \hbar^2 \mathbf{d}^2) \end{array} \right),
\end{equation}
where $\hat{\mathbf{p}} = -i \hbar \nabla$ is the nuclear momentum operator and $\mathbf{d}(\mathbf{q}) = \bra{+} \nabla \ket{-}$ is defined as the nonadiabatic coupling vector.  The adiabatic Hamiltonian can then be written as $\mathbf{H}_A = \mathbf{U}^\dagger(\mathbf{q}) \mathbf{H}_D \mathbf{U}(\mathbf{q}) = \mathbf{T}_A + \mathbf{V}_A$.  (In what follows, we will neglect the diagonal Born-Oppenheimer corrections $\hbar^2 \mathbf{d}^2/2m$.) 

The system density matrix in the diabatic representation is 
\begin{equation}
\hat{\rho}_D = \left( \begin{array}{lr}
 \hat{\rho}_{11} & \hat{\rho}_{12}\\
\hat{\rho}_{21} &  \hat{\rho}_{22} \end{array} \right),
\end{equation}
where $\hat{\rho}_{ij}$ are operators acting on the nuclear degrees of freedom.  The corresponding adiabatic density matrix is
\begin{equation}
\hat{\rho}_A = \left( \begin{array}{lr}
 \hat{\rho}_{++} & \hat{\rho}_{+-}\\
\hat{\rho}_{-+} &  \hat{\rho}_{--} \end{array} \right).
\end{equation}
The two representations are connected by the unitary transformation $\mathbf{U}(\mathbf{q})$:
\begin{equation}
\hat{\rho}_A = \mathbf{U}^\dagger \hat{\rho}_D \mathbf{U}.
\end{equation}
The quantum Liouville equation for the electronic-nuclear dynamics in the diabatic representation is
\begin{equation}
i \hbar \frac{d \hat{\rho}_D}{dt} = [\mathbf{H}_D,\hat{\rho}_D].
\end{equation}
In the adiabatic representation it is given by 
\begin{equation}
i \hbar \frac{d \hat{\rho}_A}{dt} = [\mathbf{H}_A,\hat{\rho}_A].
\end{equation}

A mixed quantum-classical description of the electronic-nuclear dynamics is obtained by representing the nuclear degrees of freedom in the Wigner-Moyal representation,\cite{wign32,zachos} an exact phase space description of quantum mechanics, and then taking the semiclassical limit.  This approximation retains only the lowest order terms in $\hbar$ of the Moyal series expansion of the exact nuclear phase space formalism. 

Following this program, the  quantum-classical Liouville equation in the diabatic representation is found to be\cite{mart97}
\begin{equation} \label{eq:scrho11}
\frac{\partial \rho_{11}}{\partial t}=\{H_{11},\rho_{11}\} + \{V,\alpha\}-\frac{2 V}{\hbar} \beta
\end{equation}
\begin{equation} \label{eq:scrho22}
\frac{\partial \rho_{22}}{\partial t}=\{H_{22}, \rho_{22}\} + \{V,\alpha \}+\frac{2 V}{\hbar}\beta
\end{equation}
\begin{equation} \label{eq:scalpha}
\frac{\partial \alpha}{\partial t}=\{H_0,\alpha\} +  \omega \beta
+\frac{1}{2}\{V,\rho_{11}+\rho_{22}\}
\end{equation}
\begin{equation} \label{eq:scbeta}
\frac{\partial \beta}{\partial t}=\{H_0,\beta\} - \omega \alpha +
\frac{V}{\hbar} \left(\rho_{11}-\rho_{22}\right).
\end{equation}
The diabatic coherence is expressed above in terms of its real and imaginary parts as $\rho_{12}= \alpha + i \beta$, and we have defined the average Hamiltonian $H_0 =(H_{11}+H_{22})/2$ and frequency $\omega=(H_{11}-H_{22})/\hbar$.  All higher order terms in $\hbar$  have been neglected, leading to a classical-limit formalism that retains only the most important nonclassical corrections.\cite{green2023zombie}

The quantum-classical Liouville equation for the system in the adiabatic representation is\cite{dono00,martens2019surface,martens2020faraday}
\begin{equation} \label{eq:adrhopp}
\frac{\partial \rho_{++}}{\partial t}=\{H_{++},\rho_{++}\} 
- \hbar \left\{ \mathbf{d} \cdot \mathbf{v}, \beta \right\} - 2 \mathbf{d} \cdot \mathbf{v} \,\alpha
\end{equation}
\begin{equation} \label{eq:adrhomm}
\frac{\partial \rho_{--}}{\partial t}=\{H_{--},\rho_{--}\} 
- \hbar \left\{ \mathbf{d} \cdot \mathbf{v}, \beta \right\} + 2 \mathbf{d} \cdot \mathbf{v} \,\alpha
\end{equation}
\begin{equation} \label{eq:adalpha}
\frac{\partial \alpha}{\partial t}=\{H_{o},\alpha \} + \omega \beta +  \mathbf{d} \cdot \mathbf{v} \,(\rho_{++} - \rho_{--})
\end{equation}
\begin{equation} \label{eq:adbeta}
\frac{\partial \beta }{\partial t}=\{H_{o},\beta\} -\omega \alpha - \frac{\hbar}{2} \left\{ \mathbf{d} \cdot \mathbf{v}, \rho_{++}+\rho_{--} \right\},
\end{equation}
where here $\rho_{+-}=\alpha + i \beta$, $H_{++} = \mathbf{p}^2/2m + V_+(\mathbf{q})$, $H_{--} = \mathbf{p}^2/2m + V_-(\mathbf{q})$, $H_o=\frac{1}{2}(H_{++}+H_{--})$, and $\omega(\mathbf{q}) = (V_+(\mathbf{q})-V_-(\mathbf{q}))/\hbar$.  $V_+(\mathbf{q})$ and $V_-(\mathbf{q})$ are the adiabatic potentials---the position-dependent eigenvalues of the diabatic potential matrix.  The quantity $\mathbf{v}(\mathbf{q},\mathbf{p})=\dot{\mathbf{q}}$ is the velocity.  It should be noted that, in the adiabatic representation, the canonical momentum $\mathbf{p}$ is not equal to $m \dot{\mathbf{q}}$.  The canonical momentum appears in the diagonal kinetic energy, while the velocity appears in nonadiabatic coupling terms. \cite{note1}   

The density matrix elements $\rho_{mn}(\mathbf{q},\mathbf{p},t)$ are now phase space Wigner functions rather than quantum operators. The relations between these functions in the diabatic and adiabatic representations are nontrivial. \cite{Martens:2016hd} In the general case, the \emph{exact} formal relationship can be obtained by employing the Moyal, or star, product\cite{zachos} rather than the simple matrix product in the unitary transformation equations.  In the semiclassical limit, an extensive discussion and explicit transformation equations are presented in Ref.~\cite{Martens:2016hd}  

For localized density matrix elements, we can simplify the transformations of Ref.~\cite{Martens:2016hd}  The resulting expressions for the adiabatic matrix elements in terms of diabatic elements become
\begin{equation} \label{eq:tpp}
\rho_{++} = \frac{\rho_{11}+\rho_{22}}{2} + \frac{\rho_{11} -\rho_{22}}{2} \cos \phi + \mathrm{Re} \, \rho_{12} \sin \phi
\end{equation}
\begin{equation}
\rho_{--} = \frac{\rho_{11}+\rho_{22}}{2} - \frac{\rho_{11} -\rho_{22}}{2} \cos \phi - \mathrm{Re} \, \rho_{12} \sin \phi
\end{equation}
\begin{equation}
\alpha = -\frac{\rho_{11}-\rho_{22}}{2}\sin \phi + \mathrm{Re} \, \rho_{12} \cos \phi 
\end{equation}
\begin{equation} \label{eq:tb}
\beta  = \mathrm{Im} \, \rho_{12},
\end{equation}
with a corresponding set of equations for the inverse transformation.\cite{Martens:2016hd}  (This is equivalent to approximating the Moyal product as the algebraic product in the transformation equations.\cite{zachos})

\section{Quantum trajectory surface hopping (QTSH)} \label{sec:sec3}

In this Section, we briefly review the quantum trajectory surface hopping (QTSH) method \cite{martens2019surface,martens2020faraday} as implemented in the adiabatic representation.  The QTSH method solves the coupled partial differential equations in Eqs.~(\ref{eq:adrhopp})--(\ref{eq:adbeta}) using a stochastic  ensemble of independent trajectories. The hopping of the trajectories is treated using the fewest switches algorithm of FSSH. \cite{tull90,martens2019surface,martens2020faraday}  In particular, for an ensemble of size $N$, each independent trajectory carries with it a ``proxy'' density matrix $a_{mn,j}$ $(j=1,2,\ldots,N)$ that represents approximately the electronic state of the system. \cite{martens2019surface,martens2020faraday}  

The QTSH treatment of the classical motion, on the other hand, differs significantly from the FSSH approach.  The trajectory equations of motion can be derived by considering the time evolution of the total nuclear density $\rho = \rho_{++} + \rho_{--}$.   Equations of motion for the trajectory variables $(\mathbf{q}_j(t),\mathbf{p}_j(t))$ ($j=1,2,\ldots,N$) are derived by demanding that the phase space partial differential equation evolves according to the quantum-classical Liouville equation. \cite{martens2019surface,martens2020faraday}   

Defining the independent variables $\Gamma = (\mathbf{q},\mathbf{p})$ and the trajectories $\Gamma_j(t) = (\mathbf{q}_j(t),\mathbf{p}_j(t))$ $(j= 1,2,\ldots,N)$, the elements of the phase space density matrix functions are each represented by the trajectory ensemble as:
\begin{equation} \label{eq:rho++}
\rho_{++}(\Gamma,t)= \frac{1}{N}\sum_{j=1}^N \sigma_j(t) \,\delta(\Gamma-\Gamma_j(t))
\end{equation}
\begin{equation} \label{eq:rho--}
\rho_{--}(\Gamma,t)= \frac{1}{N}\sum_{j=1}^N \left(1-\sigma_j(t)\right) \delta(\Gamma-\Gamma_j(t))
\end{equation}
\begin{equation} \label{eq:alpha}
\alpha(\Gamma,t)= \frac{1}{N}\sum_{j=1}^N \alpha_j(t) \delta(\Gamma-\Gamma_j(t))
\end{equation}
\begin{equation} \label{eq:beta}
\beta(\Gamma,t)= \frac{1}{N}\sum_{j=1}^N \beta_j(t) \delta(\Gamma-\Gamma_j(t)).
\end{equation}
The parameters $\sigma_j(t)$ are binary integers equal to 0 or 1 depending on whether a trajectory is on the lower or upper surface, respectively.  Population transfer is accomplished by sudden random hops of $\sigma_j$ between its two values.  These surface hops are determined stochastically using the FSSH algorithm. \cite{tull90,martens2019surface,martens2020faraday}  The parameters $\alpha_j(t)$ and $\beta_j(t)$ are real numbers that correspond to the amplitude and phase of the real and imaginary parts of the coherence carried by the $j^{th}$ trajectory. 

A proxy density matrix $a_{ij}$ $(i,j=+,-)$ is employed to represent the quantum state of each independent trajectory.  The evolution of this density matrix is given by\cite{tull90,martens2019surface,martens2020faraday}

\begin{equation} \label{eq:app}
	\dot{a}_{++,j} = -2 \mathbf{d}(\mathbf{q}_j) \cdot \mathbf{v}_j  \, \alpha_j
\end{equation}

\begin{equation} \label{eq:amm}
	\dot{a}_{--,j} = 2 \mathbf{d}(\mathbf{q}_j) \cdot \mathbf{v}_j  \, \alpha_j
\end{equation}

\begin{equation} \label{eq:alpha}
\dot{\alpha}_j = \omega(\mathbf{q}_j) \beta_j + \mathbf{d}(\mathbf{q}_j) \cdot \mathbf{v}_j \,(a_{++,j} - a_{--,j})
\end{equation}
\begin{equation} \label{eq:beta}
\dot{\beta}_j = -\omega(\mathbf{q}_j) \alpha_j.
\end{equation}
Here, we have represented the coherence in terms of real and imaginary parts: $a_{+-,j} = \alpha_j + i \beta_j$.  The continuous probabilities $a_{++,j}$ and $a_{--,j}$ are related to the stochastic integers $\sigma_j$ by the consistency condition of surface hopping: $<\sigma_j> \simeq a_{++,j}$, and their evolution determines the hopping probabilities of the stochastic algorithm \cite{tull90,martens2019surface,martens2020faraday}.

The QTSH equations of motion for $\mathbf{q}_j(t)$ and $\mathbf{p}_j(t)$ are derived by considering the partial differential equation for evolution of the total nuclear density $\rho = \rho_{++} + \rho_{--}$:
\begin{equation}
	\frac{\partial \rho}{\partial t} = \{H_{++},\rho_{++}\} + \{H_{--},\rho_{--}\} - 2 \hbar\{\mathbf{d} \cdot \mathbf{v},\beta\}.
\end{equation}
The sink and source terms of the individual densities, Eqs.~(\ref{eq:adrhopp}) and (\ref{eq:adrhomm}), cancel in the equation of motion for the total nuclear density $\rho$.  What remains are the contributing separate classical evolution of the ensembles $\rho_{++}$ and $\rho_{--}$ (the first two terms above) plus a \emph{quantum} contribution $-2 \hbar \{\mathbf{d} \cdot \mathbf{v},\beta\}$. This latter term does not change the normalization of the total density, as the phase space trace of a Poisson bracket vanishes, but nonetheless leads to nonclassical corrections to the evolution of the density---and thus the motion of the underlying trajectory ensemble.  

Inserting the trajectory ensemble ansatz into the partial differential equation for the total nuclear density leads to a set of ordinary differential equations for the canonical phase space variables of the trajectories, \cite{martens2019surface,martens2020faraday}
\begin{equation} \label{eq:qdotqtraja}
\dot{\mathbf{q}}_j = \frac{\mathbf{p}_j}{m} - 2 \hbar \beta_j \frac{\mathbf{d}(\mathbf{q}_j)}{m} 
\end{equation}
\begin{equation} \label{eq:pdotqtraja}
\dot{\mathbf{p}}_j = -\nabla V(\mathbf{q}_j,\sigma_j) + 2 \hbar  \beta_j  \left( \mathbf{v}_j \cdot \nabla \right) \mathbf{d}(\mathbf{q}_j) 
\end{equation}
for $j=1,2,\ldots,N$. Here, $V(\mathbf{q}_j,\sigma_j) = \sigma_j V_+(\mathbf{q}_j) + (1-\sigma_j) V_-(\mathbf{q}_j)$ is the instantaneous (upper or lower, depending on the value of $\sigma_j$) adiabatic potential guiding the $j^{th}$ trajectory.   

We note that the canonical coordinates $(\mathbf{q}_j,\mathbf{p}_j)$ derive from a Hamiltonian incorporating a quantum state-dependent vector potential $\mathbf{A}_j = -2 \hbar \beta_j \mathbf{d}(\mathbf{q}_j)$, and so the relationship between momentum and velocity is \emph{not} one of simple proportionality.  

Following Miller and coworkers, \cite{Cotton:2017ih} we introduce the \emph{kinematic momentum} $\mathbf{p}_k$, which corresponds to the diabatic momentum or, equivalently, to the mass times velocity:

\begin{equation} \label{eq:pkin}
	{\mathbf{p}_k}_j = m \dot{\mathbf{q}}_j = \mathbf{p}_j - 2 \hbar \beta_j \mathbf{d}(\mathbf{q}_j).
\end{equation}
We can derive the classical equations of motion in terms of the kinematic $(\mathbf{q},\mathbf{p}_k)$ rather than the canonical $(\mathbf{q},\mathbf{p})$.  Taking the time derivative of Eq.~(\ref{eq:pkin}) gives
\begin{equation}
	\dot{\mathbf{p}_k}_j = \dot{\mathbf{p}}_j - 2 \hbar \dot{\beta}_j\mathbf{d}(\mathbf{q}_j) - 2 \hbar \beta_j \dot{\mathbf{d}}(\mathbf{q}_j).
\end{equation}
We can then use the relation
\begin{equation}
 \dot{\mathbf{d}}(\mathbf{q}_j) = (\mathbf{v}_j \cdot \nabla) \, \mathbf{d}(\mathbf{q}_j)
\end{equation}
and the proxy density matrix equation of motion
\begin{equation}
\dot{\beta}_j = -\omega(\mathbf{q}_j) \alpha_j(t)
\end{equation}
to obtain, finally, the classical equations of motion for the coordinate and kinematic momentum:
\begin{equation}
\dot{\mathbf{q}}_j = \frac{{\mathbf{p}_k}_j}{m}
\end{equation}
\begin{equation}
\dot{\mathbf{p}_k}_j = -\nabla V(\mathbf{q}_j,\sigma_j) + 2 \hbar  \omega(\mathbf{q}_j) \mathbf{d}(\mathbf{q}_j) \,\alpha_j .
\end{equation}

The kinematic momentum ${\mathbf{p}_k}_j$ evolves under both the classical force $-\nabla V(\mathbf{q}_j,\sigma_j)$ and a \emph{quantum force} $ \mathbf{F}^Q_j = 2 \hbar  \omega(\mathbf{q}_j) \mathbf{d}(\mathbf{q}_j) \,\alpha_j$.  The latter depends on the nonadiabatic coupling vector $\mathbf{d}(\mathbf{q}_j)$ as well as the real part of the electronic coherence $\alpha_j$.  ($\mathbf{F}^Q_j$ can be recognized as being closely related to the off-diagonal Hellman-Feynman force. \cite{feyn39,klein1978})

The QTSH trajectories do not separately conserve energy---and there is no reason why they should!  For a quantum mechanical state, only the expectation value (as well as higher moments) of the Hamiltonian should be preserved by the time evolution.  Trajectories, in our view, are computational tools for evolving quantum \emph{states}, and should not be taken to be ``real'' physical quantities.  

The state energy $E$ in the density matrix representation corresponds to the trace $E = \mathrm{Tr} \hat{H}_A \hat{\rho}_A$.  In terms of the trajectory ensemble, this becomes
\begin{equation} \label{eq:energy}
	E = \frac{1}{N}\sum_j E_j.
\end{equation}
Due to the independent trajectory approximation, the total is an average over individual trajectory energies, each given by
\begin{equation} \label{eq:ejcanon}
	E_j = \frac{1}{2 m} \mathbf{p}^2_j + V(\mathbf{q}_j,\sigma_j) - 2 \hbar \beta_j \mathbf{d}_j(\mathbf{q}_j)\cdot \frac{{\mathbf{p}_k}_j}{m},
\end{equation}
where $V(\mathbf{q}_j,\sigma_j)=\sigma_j V_+(\mathbf{q}_j) + (1-\sigma_j) V_-(\mathbf{q}_j)$.  (Note that the canonical momentum $\mathbf{p}_j$ appears in the quadratic kinetic energy, while the kinematic momentum ${\mathbf{p}_k}_j$ appears in the coherence energy, as discussed above.)  

The $j^{th}$ trajectory energy can be written in a simpler form in terms of the kinematic momentum:
\begin{equation} \label{eq:ejkinem}
	E_j = \frac{1}{2 m} {\mathbf{p}_k}_j^2 + V(\mathbf{q}_j,\sigma_j).
\end{equation}
(Eqs.~(\ref{eq:ejcanon}) and (\ref{eq:ejkinem}) differ by a term proportional to $\hbar^2 \beta_j^2 \mathbf{d}_j^2/m$, which we ignore here. This approximation is consistent with neglecting the higher order $\hbar^2 \mathbf{d}^2/2m$ diagonal Born-Oppenheimer corrections.)

By taking the time derivative of Eq.~(\ref{eq:energy}) and employing the QTSH equations of motion, it can be shown that the total energy $E$ is conserved within the so-called consistency of surface hopping, which equates the average change of the stochastic variables $\sigma_j$ with the continuous populations of the proxy density matrices \cite{martens2019surface,martens2020faraday}:
\begin{equation}
 <\dot{\sigma}_j> \simeq \dot{a}_{++,j}.  
\end{equation}

The classical trajectory equations for the QTSH method should be contrasted with those of the fewest switches surface hopping (FSSH) approach \cite{tull90,Subotnik:2016ch,Wang:2016bz}.  In FSSH, there is no distinction between kinematic and canonical momenta. The phase space variables are treated purely classically between hops:
\begin{equation}
\dot{\mathbf{q}}_j = \frac{{\mathbf{p}_k}_j}{m}
\end{equation}
\begin{equation}
\dot{\mathbf{p}_k}_j = -\nabla V(\mathbf{q}_j,\sigma_j).\end{equation}
When a state transition occurs, the FSSH method imposes energy conservation by instantaneously changing the momentum by a \emph{momentum jump} 
\begin{equation}
{\mathbf{p}_k}_j \rightarrow {\mathbf{p}_k}_j + \Delta {\mathbf{p}_k}_j.
\end{equation}
For a hop down or up in energy, the boost in momentum is chosen to satisfy 
\begin{equation} \label{eq:hopjump}
	\frac{(\mathbf{p}_k + \Delta \mathbf{p}_k)^2}{2m} = \frac{\mathbf{p}_k^2}{2m} \pm \hbar \omega(\mathbf{q}).
\end{equation}
Where $+$ and $-$ correspond to downward and upward transitions, respectively.  For upward hops, an additional complexity occurs if the right side of Eq.~(\ref{eq:hopjump}) is negative. In that case, the classical trajectory does not have enough energy to attain the upper electronic state, and the hop is said to be ``frustrated'' and does not occur, even if the stochastic algorithm predicts it.  This discrepancy erodes the consistency of the surface hopping method.  

The treatment of energy conservation by FSSH is an \emph{ad hoc} algorithm that is based on physical reasoning rather than by any rigorous appeal to the underlying theory.  Nonetheless, the physical reasoning is sound, and in the correct limit of a temporally localized transition of a spatially localized state with well-defined initial and final momenta and kinetic energy, a rigorous theory should indeed reproduce the FSSH algorithm.  We investigate this rigorous connection in the next Section.   

\section{Derivation of FSSH momentum jumps from  QTSH} \label{sec:sec4}

We now consider a simple two state model of a localized nonadiabatic transition.  We first analyze the case where the system starts initially on the upper adiabatic surface and experiences an electronic transition to the lower adiabatic surface that is accompanied by nuclear dynamics.  We assume that the transition occurs in a localized region around, for instance, an avoided crossing, and that the population transfer is complete.  Our aim  is to investigate the energy flow between the electronic and nuclear degrees of freedom that accompanies this transition using the QTSH formalism. \cite{martens2019surface,martens2020faraday}  

The initial adiabatic density matrix for this process is of the form
\begin{equation}
\rho_A(t_o) = \left( \begin{array}{lr}
\rho(t_o) & 0\\
0 &  0 \end{array} \right)
\end{equation}
with all probability on the upper state, while the final density matrix is
\begin{equation}
\rho_A(t_f) = \left( \begin{array}{lr}
0 & 0\\
0 &  \rho(t_f) \end{array} \right),
\end{equation}
with complete population transfer to the lower state.  Here $t_o$ and $t_f$ are the initial and final times, respectively.  

We take $\rho(\mathbf{q},\mathbf{p},t)$ to be a localized state in phase space evolving under coupled electron-nuclear dynamics.  In the limit of complete localization, one can consider $\rho(\mathbf{q},\mathbf{p},t) \simeq \delta(\mathbf{q} - \mathbf{q}(t))\delta(\mathbf{p}-\mathbf{p}(t))$, in other words, a single trajectory.   

In the diabatic representation, the initial and final density matrices correspond to the same diabatic state being populated, so $\rho_D(t)$ can be written throughout the process as
\begin{equation} \label{eq:diabloc}
\rho_D(t) = \left( \begin{array}{lr}
\rho(t)  & 0\\
0 &  0 \end{array} \right).
\end{equation}
We assume that this form holds approximately for all $t \in (t_o,t_f)$.  In this limit, the adiabatic density matrix can, to a good approximation, be written as
\begin{equation}
\rho_A(t) = \left( \begin{array}{cc}
 \frac{1}{2}(1 + \cos \phi(t)) & \,\,\,\,\,-\frac{1}{2} \sin \phi(t)\\
-\frac{1}{2} \sin \phi(t) &  \,\,\,\,\, \frac{1}{2} (1 - \cos \phi(t)) \end{array} \right)\rho(t).
\end{equation}
where the angle $\phi(t)$ is evaluated at the center of the localized state $\rho(q,p,t)$.  Here, we have employed the transformation equations (\ref{eq:tpp})--(\ref{eq:tb}) and the diabatic state ansatz, Eq.~(\ref{eq:diabloc}).  


\begin{figure}
\centering
  \includegraphics[height=5cm]{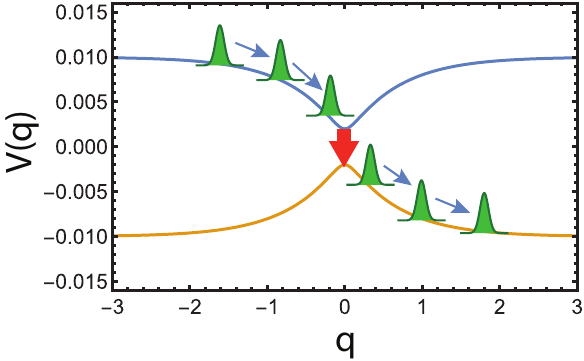}
  \caption{Schematic representation of a localized quantum state undergoing a nonadiabatic transition from the upper to lower adiabatic state at an avoided crossing. Potentials and parameters are described in the text.}
  \label{fig:fig1}
\end{figure}
\begin{figure}
\centering
  \includegraphics[height=5.5cm]{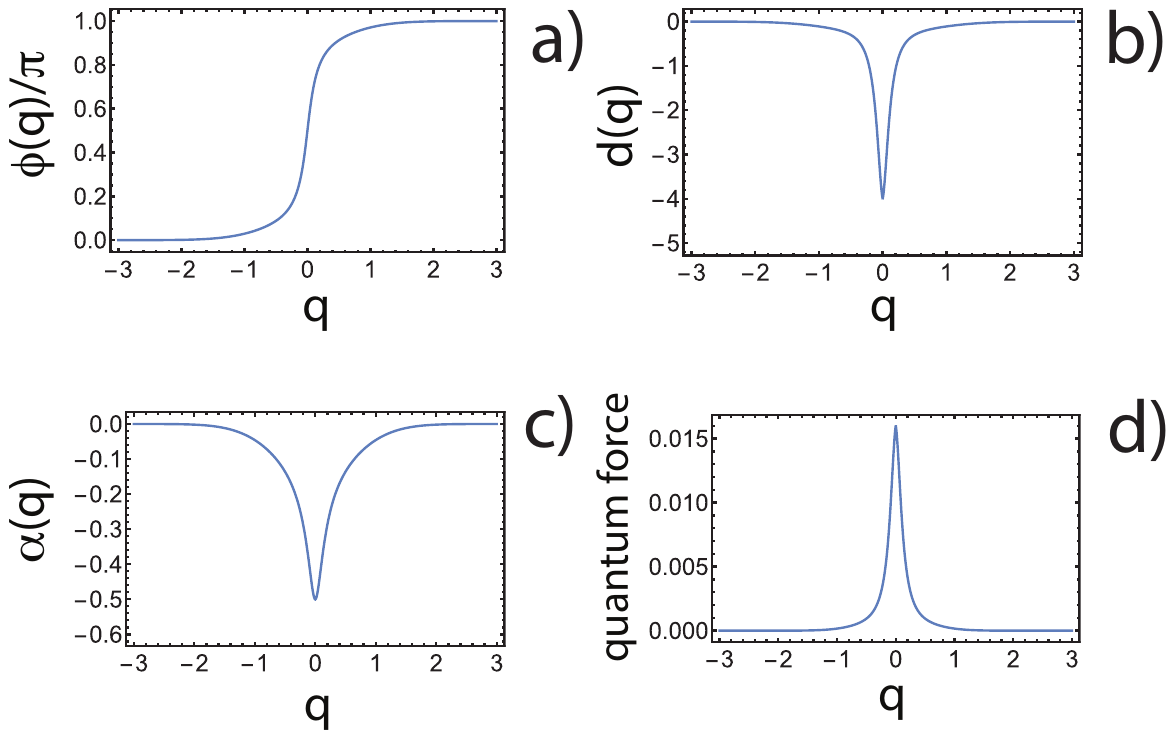}
  \caption{(a) The nonadiabatic mixing angle $\phi(q)$, (b) the nonadiabatic coupling vector $d(q)$, (c) the real part of the coherence $\alpha(q)$, and (d) the quantum force $F^Q(q)$ for the process shown in Fig.~(\ref{fig:fig1}), as described in the text.}
  \label{fig:fig2}
\end{figure}

In Fig.~(\ref{fig:fig1}) we depict schematically the nonadiabatic transition of a localized state in one dimension.  The system begins on the upper adiabatic potential and evolves until it encounters an avoided crossing with the lower adiabatic state at $q^*=0$.  At this point, the population transfers to the lower potential, indicated by the red downward arrow, where it resumes its evolution.  The motion of the state is in the positive $q$ direction, indicating that the kinematic momentum $p_k$ is positive.  During the transition, we expect the value of $p_k$ to become larger and the kinetic energy $p_k^2/2m$ to increase by a quantity equal to the energy gap $\hbar \omega(q^*)$.  The state is localized in coordinate space (and in momentum space) throughout the process.  

The potentials used to produce Fig.~(\ref{fig:fig1}) are generalizations of Tully's original system. \cite{tull90}  In particular, the adiabatic potentials are $V_1(q) = \mathrm{sgn}(q) a (1 - \exp(-b |q|)$ and $V_2(q) = -V_1(q)$. The off-diagonal diabatic coupling is $V_{12}(q) = c \exp(-d q^2)$.  The numerical values of the potential parameters, in atomic units, are $a=0.01$, $b=1.6$, $c=0.002$, and $d=1.0$.  The mass is $m=2000$ atomic units.  The value of $c$ in the present paper is smaller than the $c=0.005$ value in Ref. \cite{tull90}, leading to stronger and more localized nonadiabatic coupling in our model. 

In Fig.~(\ref{fig:fig2}) we show the nonadiabatic mixing angle $\phi(q)$ (a), the nonadiabatic coupling vector $d(q)$ (b), the real part of the coherence $\alpha(q)$ (c), and the resulting quantum force $F^Q(q) = 2 \hbar \omega(q) d(q) \alpha(q)$ (d) given by our model potentials.  The coupling, coherence, and quantum force exhibit changes in a relatively localized region around the avoided crossing, in qualitative agreement with the limit considered analytically here.  The coupling $d(q)$ and coherence $\alpha(q)$ are both negative throughout the process.  The localized quantum force, on the other hand, is a positive impulsive term, consistent with its role in increasing the momentum and kinetic energy of the nuclear motion.  

The total energy of the state of the system is given by $E = \mathrm{Tr} (\mathbf{H}_A \rho_A)$. Using the Hamiltonian and density matrix, this becomes
\begin{equation} \label{eq:ecanon}
	E(t) = \frac{\mathbf{p}^2}{2 m} + V(\mathbf{q},\sigma) - 2 \hbar \beta \mathbf{d}(\mathbf{q})\cdot \frac{{\mathbf{p}_k}}{m},
\end{equation} 
where $(\mathbf{q}(t),\mathbf{p}(t))$ is our trajectory and $V(\mathbf{q},\sigma) = \sigma V_+(\mathbf{q}) + (1-\sigma) V_-(\mathbf{q})$. (Our ensemble here has reduced to a single trajectory, and so we have dropped the subscript $j$).  The canonical classical energy in the first two terms is augmented by a \emph{coherence energy} $E_{coh.}= - 2 \hbar \beta \mathbf{d}(\mathbf{q}(t))\cdot \frac{{\mathbf{p}_k}(t)}{m}$.  

As noted above, the energy can be written more simply in terms of the kinematic momentum $\mathbf{p}_k (t)= \mathbf{p}(t)-2 \hbar \beta (t) \mathbf{d}(\mathbf{q} (t))$.  We find
\begin{equation} \label{eq:ekinem}
	E = \frac{\mathbf{p}_k^2}{2 m} + V(\mathbf{q},\sigma)=\frac{\mathbf{p}_k^2}{2 m} + \sigma \hbar \omega(\mathbf{q}) + V_-(\mathbf{q}),
\end{equation}
where $\hbar \omega(\mathbf{q}) = V_+(\mathbf{q}) - V_-(\mathbf{q})$.  

Using the kinematic momentum, we can separate the energy into two terms: the electronic state-dependent potential energy $E_{elec.}=V(\mathbf{q},\sigma)$ and the classical kinetic energy $E_{kin.}=\mathbf{p}_k^2/2 m$ that depends only on the nuclear motion. In this representation, the coherence energy does not appear in the energy budget. Quantum effects are nonetheless still present.  

In the Born-Oppenheimer approximation, the electronic parameter $\sigma(t)$ is constant, assuming values of 0 or 1.  In this limit, the nuclear dynamics are governed by purely \emph{classical} motion on the appropriate adiabatic potential, $V_+$ or $V_-$ for $\sigma=1$ or $\sigma=0$, respectively.  The equations of motion for the classical variables and electronic state parameters in the Born-Oppenheimer limit are 
\begin{equation}
	\dot{\mathbf{q}} = \frac{\mathbf{p}_k}{m}
\end{equation}
\begin{equation}
	\dot{\mathbf{p}_k} = -\nabla V(\mathbf{q},\sigma)
\end{equation}
\begin{equation}
	\dot{\sigma} = 0
\end{equation}
\begin{equation}
	\dot{\alpha} = \omega(\mathbf{q}) \beta
\end{equation}
\begin{equation}
	\dot{\beta} = - \omega(\mathbf{q}) \alpha.
\end{equation}
When nonadiabatic transitions occur, however, \emph{nonclassical} forces appear, as described by the QTSH formalism.  The equations of motion are generalized to
\begin{equation}
	\dot{\mathbf{q}} = \frac{\mathbf{p}_k}{m}
\end{equation}
\begin{equation}
	\dot{\mathbf{p}_k} = -\nabla V(\mathbf{q},\sigma) + 2 \hbar \omega(\mathbf{q}) \mathbf{d}(\mathbf{q}) \,\alpha 
\end{equation}
\begin{equation} \label{eq:dotsigma}
	\dot{\sigma} = -2 \frac{\mathbf{d}(\mathbf{q}) \cdot \mathbf{p}_k}{m} \alpha
\end{equation}
\begin{equation}
	\dot{\alpha} = \omega(\mathbf{q}) \beta +\frac{\mathbf{d}(\mathbf{q}) \cdot \mathbf{p}_k}{m} (2 \sigma - 1)
\end{equation}
\begin{equation}
	\dot{\beta} = - \omega(\mathbf{q}) \alpha
\end{equation}
Here, we have made the substitution of the stochastic integer parameter $\sigma$ for the continuous electronic population $a_{++}$.  The consistency of surface hopping is equivalent to assuming that the average $\sigma$ is equal to the continuous population parameter across a trajectory ensemble: $<\sigma>\simeq a_{++}$.  For localized transitions resulting in complete population transfer, both of these quantities are integers except during a brief excursion, where their values change by unity.  

Such a transition occurs at a localized crossing time $t = t^*$, or, equivalently, around a localized configuration $\mathbf{q}(t^*) = \mathbf{q}^*$.   We shall calculate the changes in the constituents of the total energy as well as the work done by the quantum forces that the electronic and nuclear degrees of freedom exert on each other during that transition.  

The electronic transition from the upper ($\sigma = 1$) state to the lower ($\sigma = 0$) state is assumed to be localized within a short time interval of duration $2 \epsilon$ that is symmetric around $t = t^*$.  During this interval, we assume that the classical forces and dynamics leave $\mathbf{q}$ and $\mathbf{p}_k$ essentially unchanged.  The change in $\sigma$ during this interval is given by
\begin{equation}
	\Delta \sigma = -1.
\end{equation}

For a fixed nuclear coordinate $\mathbf{q} = \mathbf{q}^*$, the change in electronic energy during the transition is
\begin{equation}
	\Delta E_{elec.} = \hbar \omega(\mathbf{q}^*) \Delta \sigma = -\hbar \omega(\mathbf{q}^*).
\end{equation}
This change in electronic energy is negative for a transition from upper to lower state.  To conserve the total energy, the classical kinetic energy must increase by the same amount:
\begin{equation}
	\Delta E_{kin.} = -\Delta E_{elec.} = +\hbar \omega(\mathbf{q}^*).
\end{equation}
The FSSH method imposes this conservation in an \emph{ad hoc} manner by artificially rescaling the momentum by a ``jump'', chosen to satisfy 
\begin{equation} \label{eq:tullyquadratic}
	\frac{(\mathbf{p}_k + \Delta \mathbf{p}_k)^2}{2m} = \frac{\mathbf{p}_k^2}{2m} + \hbar \omega(\mathbf{q}).
\end{equation}
The momentum jump $\Delta \mathbf{p}_k$ is chosen to solve this quadratic equation in a direction parallel to the nonadiabatic coupling vector $\mathbf{d}(\mathbf{q}^*)$. Practical implementations have rules for selecting which root to choose, and what to do if no solution can be found due to insufficient energy or directional constraints (a ``frustrated hop'').\cite{Subotnik:2016ch,Wang:2016bz} 

We now analyze the energy budget from the QTSH perspective.  From the equations of motion, we have
\begin{equation}
	\dot{\mathbf{p}}_k = -\nabla V(\mathbf{q},\sigma) + 2 \hbar \omega(\mathbf{q}) \mathbf{d}(\mathbf{q}) \alpha.
\end{equation}
The change in the kinematic momentum results from two contributions: the classical force derived from the currently occupied adiabatic potential and a quantum force resulting from the electronic energy transition.  This change of electronic states does work on the nuclear degrees of freedom.  We will calculate the result of this work in what follows.  

During the localized transition the classical force does not have an appreciable effect on the nuclear dynamics.  We consider only the impulsive quantum force during the transition.  Here, we can take the coordinate $\mathbf{q}=\mathbf{q}^*$ to be constant.  The integrated effect of this force on the momentum during this transition can be computed as
\begin{equation}
	\Delta \mathbf{p}_k = 2 \hbar \omega(\mathbf{q^*}) \mathbf{d}(\mathbf{q}^*)\int_{t^*-\epsilon}^{t^*+\epsilon} \alpha(t) dt.
\end{equation}
The only time-dependent quantity during the transition is the electronic coherence, which is rapidly created and then dispersed by the flow of electronic population from upper to lower state.  

We showed above that this coherence can be described simply in terms of the nonadiabatic mixing angle:
\begin{equation}
	\alpha(t) = -\frac{1}{2} \sin \phi(t)
\end{equation}
The value of $\phi$ transitions rapidly from $\phi=0$ for $t<t^*$ to $\phi = \pi$ for $t>t^*$.  To simplify the integral, we change integration variables from $t$ to $\phi$.  We note that 
\begin{equation}
d \phi = \dot{\phi} \, dt.
\end{equation}
We can therefore write $\dot{\phi}$ as
\begin{equation}
	\dot{\phi} = \nabla \phi \cdot \dot{\mathbf{q}} = \nabla \phi \cdot \frac{\mathbf{p}_k}{m}.
\end{equation}
Using the definition of the nonadiabatic coupling vector
\begin{equation}
	\mathbf{d} = -\frac{1}{2} \nabla \phi
\end{equation}
gives
\begin{equation}
	dt = -\left(\frac{m}{2 \,\mathbf{d}(\mathbf{q}^*) \cdot \mathbf{p}_k}\right) d\phi.
\end{equation}
The momentum jump can then be calculated:
\begin{equation}
	\Delta \mathbf{p}_k = \frac{1}{2} \hbar \omega(\mathbf{q^*}) \mathbf{d}(\mathbf{q}^*) \left(\frac{m}{\mathbf{d}(\mathbf{q}^*) \cdot \mathbf{p}_k}\right)\int_0^\pi \sin(\phi) d\phi.
\end{equation}
Noting that $\int_0^\pi \sin(\phi) d\phi = 2$, we obtain the final result:
\begin{equation} \label{eq:qtshjump}
	\Delta \mathbf{p}_k = \hbar \omega(\mathbf{q^*}) \mathbf{d}(\mathbf{q}^*) \left(\frac{m}{\mathbf{d}(\mathbf{q}^*) \cdot \mathbf{p}_k}\right).
\end{equation}

From the perspective of the energy budget, the quantum $\mathbf{F}^Q$ does work $W_{e \rightarrow n}$ on the nuclear degrees of freedom, which changes the kinetic energy by an amount $\Delta E_{kin.} \cite{gold80}$:
\begin{equation}
W_{e \rightarrow n} = \Delta E_{kin.} = \int_{t_o}^{t_f} \mathbf{F}^Q(t) \cdot \mathbf{v}(t) \, dt
\end{equation}
In our localized approximation, this becomes
\begin{equation}
	W_{e \rightarrow n} = \Delta E_{kin.} = \frac{\mathbf{p}_k}{m} \cdot \Delta \mathbf{p}_k=\hbar \omega(\mathbf{q}^*).
\end{equation}
We recover the energy conservation expected on physical grounds: the electronic energy decrease $\hbar \omega(\mathbf{q}^*)$ that accompanies the downward electronic transition appears as kinetic energy of the classical nuclear motion.  The quantum force $\mathbf{F}^Q$ that accompanies the loss of electronic energy does work on the nuclear degrees of freedom and quantitatively transfers this energy into their kinetic energy. 

By using the equation of motion for $\sigma(t)$, Eq.~(\ref{eq:dotsigma}), it is straightforward to show that $W_{n \rightarrow e}$, the work done \emph{on} the electronic degrees of freedom \emph{by} the nuclear degrees of freedom, obeys $W_{n \rightarrow e} + W_{e \rightarrow n} = 0$, ensuring that the total energy is conserved.  

We note that the explicit expression for the momentum jump, Eq.~(\ref{eq:qtshjump}) resolves the ambiguity of the quadratic equation root choice inherent in the FSSH methodology. \cite{tull90,Subotnik:2016ch,Wang:2016bz}

We now consider the case of nonadiabatic transitions from a lower to an upper energy electronic state.  The analysis is similar, but with the boundary conditions of the angle $\phi$ being $\phi= \pi$ for $t <t^*$ and $\phi = 0$ for $t > t*$.  We find that the momentum jump for the upward transitions is the negative of the previous result:
\begin{equation}
	\Delta \mathbf{p}_k = \frac{1}{2} \hbar \omega(\mathbf{q^*}) \mathbf{d}(\mathbf{q}^*) \left(\frac{m}{\mathbf{d}(\mathbf{q}^*) \cdot \mathbf{p}_k}\right)\int_\pi^0 \sin(\phi) d\phi
\end{equation}
or
\begin{equation} \label{eq:upjump}
	\Delta \mathbf{p}_k = - \hbar \omega(\mathbf{q^*}) \mathbf{d}(\mathbf{q}^*) \left(\frac{m}{\mathbf{d}(\mathbf{q}^*) \cdot \mathbf{p}_k}\right).
\end{equation}

The additional consideration of available energy enters the analysis for upward transitions. In particular, enough nuclear kinetic energy must be available to make the transition to the upper state.  This corresponds to the positive kinetic energy before the transition $\mathbf{p}_k^2/2m$ to be at least as large in magnitude as the \emph{negative} $\Delta E_{kin.}$ accompanying the electronic excitation.  Otherwise the hop is ``frustrated'', as previously discussed.  

We can examine frustrated hops in the context of the present analysis. In order for $\phi$ to fully transition from $\phi=\pi$ to $\phi=0$ there must be enough kinetic energy to keep the nuclear motion proceeding in its initial positive direction.  If this is not the case, the sign of $\mathbf{p}_k$ will reverse before complete population transfer occurs, reversing the direction of motion.  Accompanying this will be a change in the sign of $\dot{\phi}$, so that $\phi$ has the time history $\pi \rightarrow \phi_{min.} \rightarrow \pi$ rather than $\pi \rightarrow 0$ for a successful upward transition, where $\phi_{min.}$ is the value reached by the mixing angle when the momentum reversal occurs.  The total work done by the quantum force in this case is $\Delta E_{kin.}=0$, the expected accompaniment for the $\Delta \sigma=0$ failed transition.  Correspondingly, the electronic degrees of freedom do zero net work on the nuclear motion for frustrated hops.  

\section{Numerical Results} \label{sec:sec5}

We now present numerical simulations of the model system given above to illustrate the general behavior of the QTSH method, and interpret the results from the perspective of our analytic results. 

\begin{figure}
\centering
  \includegraphics[height=5cm]{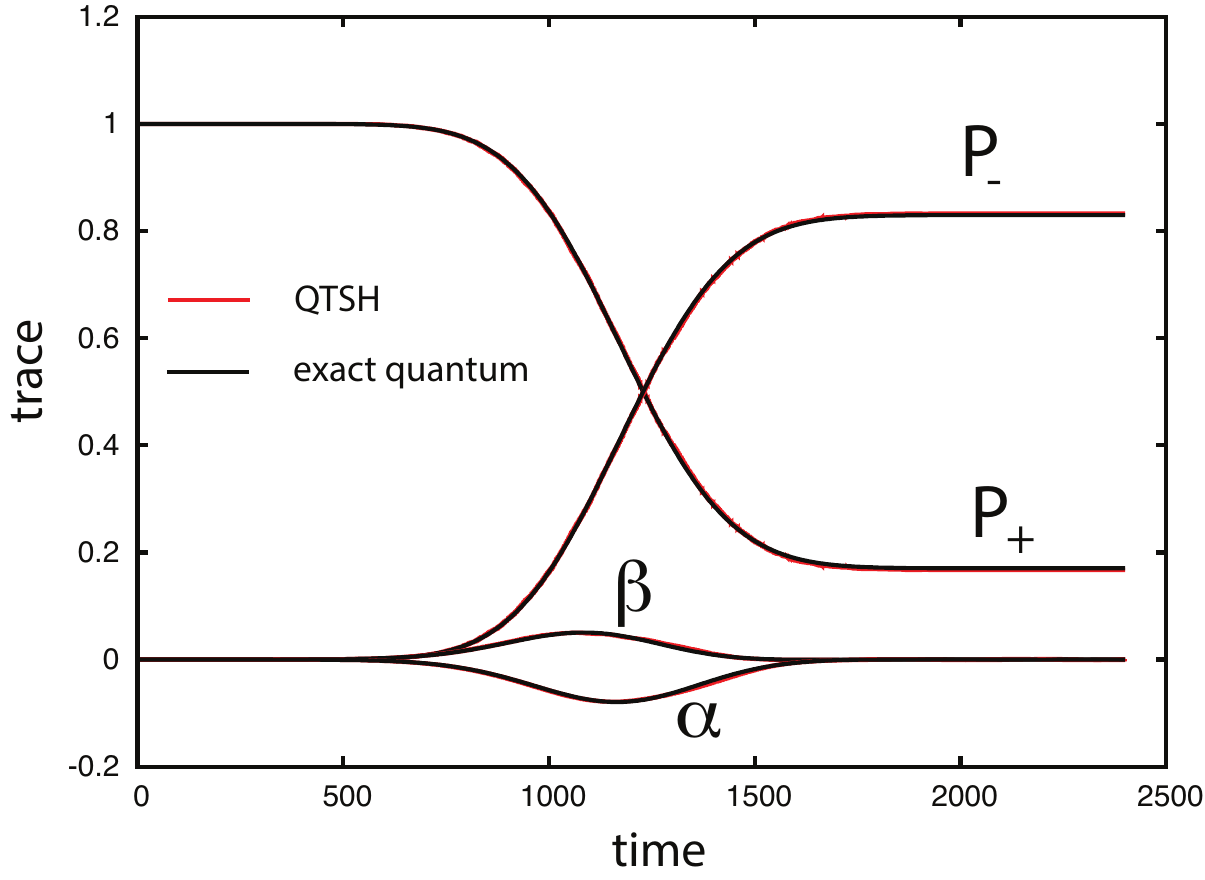}
  \caption{Comparison of QTSH simulations with exact quantum wavepacket dynamics for the system shown in Fig.~(\ref{fig:fig1}), as described in the text.}
  \label{fig:fig3}
\end{figure}

 In Fig.~(\ref{fig:fig3}) we show numerical results for our modified Tully 1 system, described above.  We show the results of a QTSH simulation with 10000 trajectories and compare our results with exact wavepacket simulations performed using the method of Kosloff \cite{kosl94}. The inital Gaussian wavepacket has a mean momentum $\hbar k = 10$.  Details of the simulation can be found in Ref. \cite{martens2019surface}  We note that here we are using the kinematic momentum rather than the canonical momentum in nonadiabatic coupling terms, which improves the accuracy of the QTSH method. \cite{note1} 
 
 The quantities shown correspond to the populations on state $\ket{+}$ and $\ket{-}$.  These are given by the trace of the corresponding density matrix elements: $P_{+} = (1/N) \sum_j \sigma_j(t)$ and $P_-(t) = 1-P_+(t)$. The quantities $\alpha(t)$ and $\beta(t)$ are the real and imaginary parts of trace of the coherence $\rho_{+-}(t)$: $\alpha(t) = (1/N)\sum_j \alpha_j(t)$ and $\beta(t) = (1/N)\sum_j \beta_j(t)$.  The quantum calculations were performed in the diabatic representation and transformed to the adiabatic representation for analysis.  For this system and initial conditions, the agreement between QTSH and exact quantum results is essentially quantitative.  
 
 \begin{figure}
\centering
  \includegraphics[height=5.2cm]{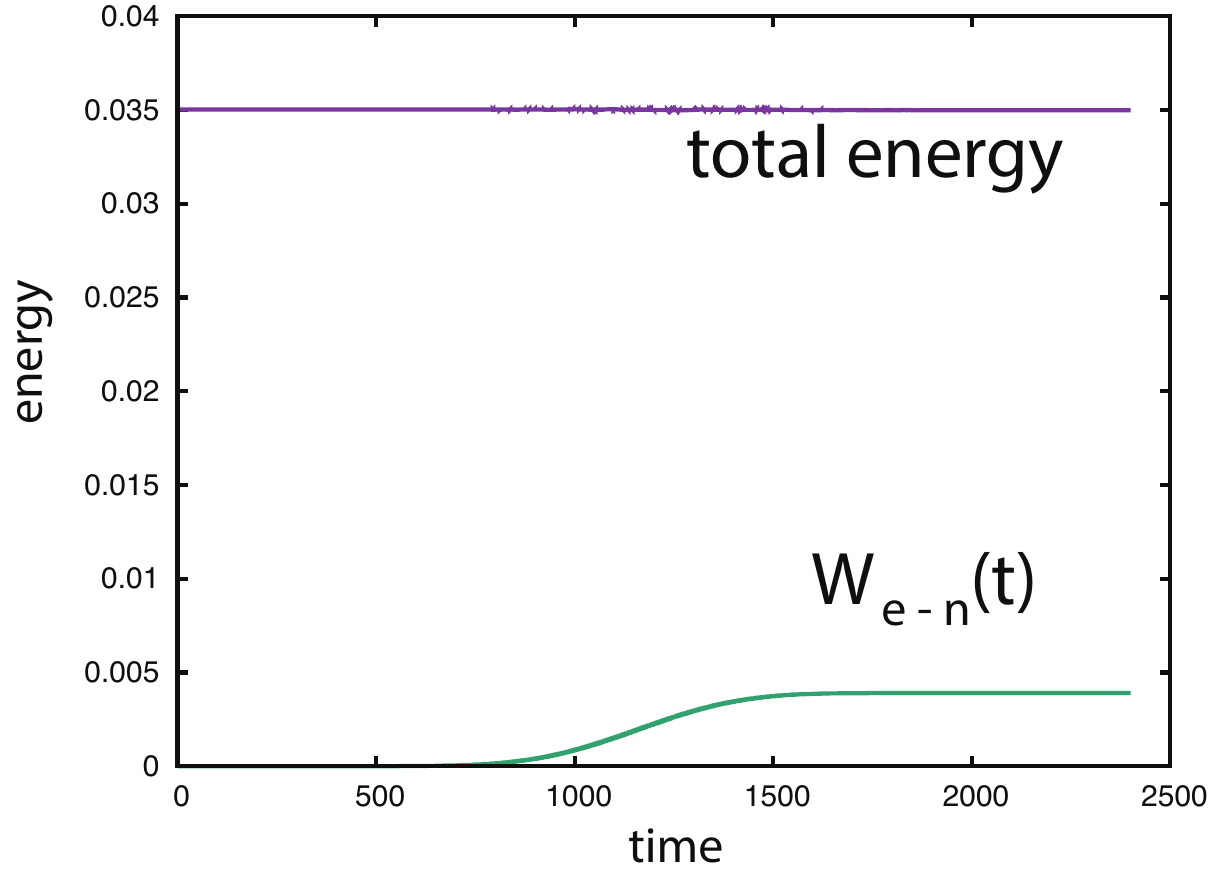}
  \caption{Total ensemble energy of the QTSH simulation shown in Fig.~(\ref{fig:fig3}).  Also shown is the accumulated work performed on the nuclear degrees of freedom by the quantum force, as discussed in the text.}.  
  \label{fig:fig4}
\end{figure}

In Fig.~(\ref{fig:fig4}) we show the total ensemble energy $E(t) = (1/N) \sum_j {p_k}_j^2 /2m + V(q_j,\sigma_j)$.  The results show that the total ensemble energy is conserved nearly exactly, indicating that energy conservation of the state of the system is indeed emerging from the quantum forces of the QTSH method \emph{without momentum rescaling}.  Individual trajectories do not separately conserve energy, but the observable energy of the state is nonetheless constant in time.  

Also shown in the Figure is the time-dependent accumulated work $W_{e \rightarrow n}(t) = (1/N) \sum_j \int_0^t F^Q_j ({p_k}_j/m) dt$.  Unlike for the impulsive analytic results, the energy exchange between electronic and nuclear degrees of freedom is spread over \emph{ca.}~500 atomic time units rather than occurring at a single point in time.  As expected, the total work done $W_{e \rightarrow n}(t_f)$ is approximately equal to the energy gap, which for our model is  $\hbar \omega(q^*)=0.004$ atomic units. 

These results demonstrate the essentially quantitative accuracy of the QTSH method for this simple and direct classically-allowed electronic relaxation.  We note that, for this system and initial condition, FSSH gives results that are essentially indistinguishable from the QTSH observables (data not shown).  Differences occur between the trajectory approaches when processes are more nonclassical, and the more ``quantum'' treatment of the energy budget allows QTSH to simulate such processes where the strict classical energy conservation of FSSH leads to significant errors. \cite{martens2020faraday}   

\section{Discussion} \label{sec:sec6}

In this paper, we have investigated the energetics of mixed quantum-classical systems from the perspective provided by quantum trajectory surface hopping (QTSH), a method for simulating molecular dynamics with electronic transitions using ensembles of independent trajectories.  In contrast with fewest switches surface hopping (FSSH), the QTSH approach does not impose energy conservation externally through instantaneous momentum jumps, but instead relies on rigorously-derived quantum forces that act continuously to establish energy conservation at the ensemble level.  

Individual QTSH trajectories do not separately conserve energy---nor should they.  Indeed, we have long emphasized that quantum effects show up naturally in trajectory ensemble methods as a relaxation of strict classical constraints on individual trajectories. \cite{dono01,dono03,riga04,riga06a,Martens:2016ea}  Independent individual trajectories are not a part of quantum theory, and should be viewed as a calculational tool rather than real physical quantities.  From the perspective of the foundations of quantum mechanics, trajectories are \emph{hidden variables}, and as Bell's theorem established, \cite{bellbook} a faithful hidden variable theory must be \emph{nonlocal}.  This nonlocality shows up in QTSH as relaxation of individual trajectory energy conservation.  

In the limit of localized nonadiabatic transitions, where the the physical assumptions behind the FSSH algorithm become quantitatively valid, the momentum jumps should emerge from an exact theory.  We establish this connection by deriving the FSSH algorithm from the QTSH equations of motion.  The analysis also resolved the ambiguity resulting from the \emph{ad hoc} root search algorithm used by FSSH to determine jumps by giving an explicit expression for the momentum rescaling, Eq.~(\ref{eq:qtshjump}).  

For general nonadiabatic transitions that are not localized in time and space, the FSSH momentum jumps are no longer rigorously valid.  The feedback between nuclear and electronic degrees of freedom is always at work, mediated by the quantum forces described here.  QTSH provides the correct treatment of the energy budget in the general case, giving nearly quantitative agreement with exact quantum simulations.  The quantum forces characterizing the QTSH method rigorously lead to energy conservation within the consistency of the underlying stochastic surface hopping without artificial external momentum rescaling.   

\section{Acknowledgements}

We thank Greg Ezra, Leonardo Araujo, and Burkhard Schmidt for helpful input and acknowledge stimulating discussions at the UCI Liquid Theory Lunch (LTL) and the Telluride Science Research Center (TSRC).  This paper builds upon work supported by the National Science Foundation under grant CHE-1764209.

\bibliography{ccm_2023}
\end{document}